\documentclass[12pt,preprint]{aastex}

\newcommand{\e}{\epsilon}

\newcommand{\Bmug}{B_{\mu{\rm G}}}
\newcommand{\KsC}{K_{syn{\rm C}}}

\shorttitle{Chandra X-ray Jets}
\shortauthors{Dermer \& Atoyan}

\begin{document}

\title{X-ray Synchrotron Spectral Hardenings from Compton and \\
 Synchrotron Losses in Extended Chandra Jets 
}

\author{Charles D.\ Dermer\altaffilmark{1} and Armen M.\ Atoyan\altaffilmark{2}}
\altaffiltext{1}{E. O. Hulburt Center for Space Research, Code 7653,
Naval Research Laboratory, Washington, DC 20375-5352}

\altaffiltext{2}{CRM, Universite de Montreal, Montreal H3C 3J7, Canada}

\begin{abstract}
{\it Chandra} observations of knots and hot spots in spatially
resolved X-ray jets of radio galaxies show that the X-ray fluxes often
lie above an extrapolation from the radio-to-optical continuum fluxes.
  We show that combined synchrotron and Compton
losses on a single power-law electron injection function can produce a
hardening in the electron spectrum at electron Lorentz factors $\gamma
\approx 2\times 10^8/[\Gamma(1+z)]$ due to KN energy losses on the cosmic
microwave background radiation. Here $\Gamma$ is the bulk Lorentz
factor of the outflow, and $z$ is the source redshift. This produces a
flattening in the spectrum at frequencies $\gtrsim 8\times 10^{16}
\delta B_{\mu{\rm G}}/[\Gamma^2(1+z)^3]$ Hz, where $B_{\mu{\rm G}}$ is
the magnetic field in the comoving plasma frame in units of
micro-Gauss and $\delta$ is the Doppler factor. A single population of
synchrotron-emitting electrons may therefore produce the
radio-to-X-ray continuum in some radio galaxy knots, such as those in
3C 273.
\end{abstract}


\section{Introduction}

The excellent spatial resolution of the {\it Chandra X-ray
Observatory} is providing X-ray images of extended radio galaxy jets
in the 0.2-8 keV band at resolutions better than $0.4^{\prime\prime}$
FWHM, and spectral detail of discrete components at flux levels between
$\approx 10^{-13}$ and $10^{-14}$ ergs cm$^{-2}$ s$^{-1}$. Combined
with high resolution radio and optical maps, spectral energy
distributions (SEDs) of spatially-resolved hot spots and knots in the
jets of radio galaxies can be studied from the radio through the X-ray
regime.

The broadband SEDs in many of the hot spots and
knots show a general behavior whereby the X-ray spectra are harder
than the optical spectra, and the X-ray fluxes are above an
extrapolation from the optical fluxes. This behavior is observed in
the SEDs of at least three of the four regions along the X-ray jet of
3C 273 \citep{sam01,mar01}, the western hot spot of Pictor A
\citep{wys01}, knot WK7.8 of PKS 0637-752
\citep{sch00,cha00}, hot spot A of Cygnus A \citep{wys00},
and in some of the knots in the jet of M87 \citep{wy01}.  In other
cases, such as the X-ray jet in 3C 66B \citep{hbw01}, knots A1 and A3
in 3C 273 (\citet{mar01}; see, however, \citet{sam01} for a different
spectral analysis), and some other knots in M87's jet \citep{wy01},
the X-ray--optical--radio spectrum is consistent with a single or
smoothly broken power-law spectrum, indicating that the broadband
emission is entirely due to nonthermal synchrotron radiation.

Three main leptonic processes have been considered to account for the
X-ray emission in the extended jets \citep{hk02}: nonthermal
synchrotron radiation, synchrotron self-Compton (SSC) radiation, and
X-ray emission from Compton-scattered external radiation
fields. Nonthermal synchrotron emission is favored for smooth spectra
and for the western hot spot of Pictor A \citep{wys01}, where an SSC
origin implies a very weak magnetic field, and a combined
SSC/synchrotron origin requires that the two emission components be
carefully matched in the X-ray regime.  The SSC process can, however, explain
the X-ray emission mechanism from the bright radio hot spots of Cygnus
A \citep{wys00}, the radio hot spots of 3C 295
\citep{har00}, and the eastern hot spot of 3C 123 \citep{hbw01a}.

\citet{tav00} and \citet{cgc01} argue that the
X-ray emission from knot WK7.8 is due to Compton-scattered cosmic
microwave background radiation (CMBR) rather than to SSC emission,
because an SSC model requires a system well out of equilibrium and a
jet that is significantly debeamed.  This model invokes an emitting
region in bulk relativistic motion on size scales of several hundred
kpc, in accord with observations of one-sidedness in the jets on these
size scales. Such a model has been applied to X-ray emitting
knots in the jet of 3C 273 by \citet{sam01}. A synchrotron model for
the X-ray emission is discounted because it would apparently require
two populations of relativistic electrons to explain the radio,
optical, and X-ray fluxes.
Another possibility is that the X-rays are 
synchrotron radiation of high-energy protons \citep{fa01}.

Here we show that synchrotron emission from a single population of
electrons injected with a nonthermal power law spectrum can produce
X-ray spectral hardenings. When CMBR cooling in the Thomson regime
exceeds synchrotron cooling, a hardening in the electron spectrum is
formed at electron energies where KN effects become important. This
produces a hardening in the synchrotron spectrum between optical and
X-ray frequencies.  In the next section, an analytic model for the
synchrotron origin of X-ray emission from extended jets is presented,
motivated by Chandra observations of 3C 273. Results of a numerical
simulation with cascade reprocessing are given in Section 3.  Section
4 gives a discussion and summary.

\section{Analytic Model }

In 3C 273, the X-ray energy spectral index in the inner knot region A
has $\alpha_x\sim 1.1$, a value characteristic of a cooling
shock-accelerated electron spectrum, and there is an indication of
spectral softening along the jet \citep{sam01}.  The peaks of the
X-ray emission along the jet of 3C 273 closely track the peaks in the
optical waveband, and the overall X-ray emission profile displays a
decreasing mean intensity along the jet, whereas the optical intensity
is roughly uniform along the jet.  The radio profile also displays
brightenings in spatial coincidence with the X-ray and optical peaks,
but the radio intensity increases dramatically along the jet. The
optical emission profiles are narrower than those of the radio
emission.

These behaviors suggest that the electron population producing the
radio emission accumulates due to weak cooling, whereas the electrons
producing the optical and X-ray emission strongly cool. Electrons that
produce synchrotron radio emission at observer's frame frequencies
$\nu$ have Lorentz factors $\gamma \approx 5\times 10^4 [(\nu/5{\rm
~GHz})(1+z)/\delta
\Bmug]^{1/2}$, whereas the electrons that Compton scatter the CMBR to
X-ray energies $E_x$ have $\gamma \approx 1\times 10^3 [(E_x/1 {\rm
~keV})/\delta\Gamma]^{1/2}$.  Consequently the X-ray emitting
electrons will cool more slowly than the radio-emitting electrons
unless $B$ exceeds milli-Gauss intensities. The fast-cooling
requirement for the X-ray emitting electrons is readily satisfied
assuming a synchrotron origin for the X-rays. In this paper, we
neglect complications that involve variations of the magnetic field
$B$ and the Doppler factor $\delta =
[\Gamma(1-\beta_\Gamma\cos\theta)]^{-1}$ along the length of the jet.
(Here $\beta_\Gamma = (1-\Gamma^{-2})^{-1/2}$ and $\theta$ is the
angle between the directions of the jet and the observer.)

The synchrotron energy-loss rate of relativistic nonthermal electrons with a
random pitch-angle distribution in a
region with comoving mean intensity $B$ is $-\dot\gamma_{syn} \cong
1.3\times 10^{-21}\Bmug^2\gamma^2$ s$^{-1}$ $\equiv k_{syn}\gamma^2$
\citep{bg70}.  Under the same conditions, the energy loss rate due to
Compton-scattered CMB radiation in the Thomson limit is
$ -\dot\gamma_{\rm T}$  = $(4c\sigma_{\rm T}/
3 m_ec^2)\gamma^2(1+z)^4 \hat u_{CMB}\Gamma^2(1+\beta_\Gamma^2/3)
$ $\cong 1.7\times 10^{-20}\gamma^2(1+z)^4\Gamma^2\;{\rm s}^{-1}\; $ $
\equiv k_{\rm T}\gamma^2 $,
where $\hat u_{CMB}\cong 4\times 10^{-13}$ ergs cm$^{-3}$ is the local
($z=0$) intensity of the CMBR. This expression holds when
$\gamma\ll \gamma_{\rm KN}$, where $\gamma_{\rm KN}$ is the Lorentz
factor above which KN effects become important, given through
$4\gamma_{\rm KN}\hat\epsilon^\prime = 1$. For the CMB radiation, the
dimensionless mean photon energy $\hat\epsilon^\prime \cong
\hat\epsilon_{CMB}\Gamma(1+z) \cong 2.7\, (k_{\rm B}\times 2.72 {\rm
K}) \Gamma (1+z)/m_ec^2 $
and $\hat\epsilon_{CMB}= 1.24\times 10^{-9}$, so that electrons with 
$\gamma \ll \gamma_{\rm KN} = 2\times 10^8/\Gamma(1+z)$ scatter CMB 
radiation in the Thomson regime. 

In the extreme KN limit, the electron energy loss rate 
$-\dot\gamma_{\rm KN}  
 \cong 4.6\times 10^{-3} (1+z)^2\ln[2.54\gamma_{10}\Gamma(1+z)]$
s$^{-1}$ \citep{bg70}, where
$\gamma_{10} = \gamma/10^{10}$,
and $\Theta = k_{\rm B}(2.72)(1+z)/m_ec^2 = 4.6\times 10^{-10}(1+z)$ is 
the dimensionless temperature. We approximate 
the combined energy-loss rate for analytic simplicity by the expression
\begin{equation}
-\dot\gamma = [k_{syn} + {k_{\rm T}\over 1+(a\gamma/\gamma_{\rm KN})^b}]
\gamma^2\;\equiv \KsC(\gamma) \gamma^2,
\label{dotgamma}
\end{equation}
where $a = 0.5$ and $b = 1.7$ 
 are chosen to fit roughly the $-\dot
\gamma\propto \ln\gamma$ behavior in the extreme KN regime and to
provide spectral results within a factor $\sim 2$-3 of the numerical
results.  Equation (1) 
shows that when $k_{\rm T} \gg k_{syn}$, $\dot
\gamma$ is dominated by Thomson losses at $\gamma \ll \gamma_{\rm
KN}$, $\dot\gamma$ flattens at $\gamma_{\rm KN}\lesssim \gamma
\lesssim \bar\gamma = \gamma_{KN}\sqrt{k_{\rm T}/k_{syn}}\equiv 
\gamma_{KN}\sqrt{N_{Ts}}$, and $\dot\gamma$
 is dominated by synchrotron losses at
$\gamma\gg\bar\gamma$. Adiabatic losses are neglected
 here but can be shown
to be negligible for X-ray emitting electrons.

For numerical calculations in the next section, we solve the
time-dependent continuity equation for the energy spectrum of
electrons $N_{e}(\gamma,t)$ injected in the source with differential
injection rate $Q(\gamma,t)$. For the analytic model treated here, we
approximate the injection rate as a stationary function starting from
$t_0=0$, with a single power-law behavior $Q(\gamma)= Q_0 \,
\gamma^{-p}\exp(-\gamma/\gamma_{max})$ for $\gamma \geq \gamma_{min}$,
implying a comoving 
frame electron power $L_e = m_ec^2
\int_{\gamma_{min}}^{\gamma_{max}} d\gamma\; \gamma Q(\gamma)$.
In the stationary injection case with $p > 1$, 
the energy distribution of 
electrons at time $t$ can be approximated by $N_{e}(\gamma,t)
\simeq Q(\gamma) t_{\rm ac} $, where the characteristic electron
accumulation time $t_{\rm ac} = {\rm min} (t,
t_{cool}=-\gamma/\dot{\gamma})$. In this approximation
\begin{equation}
\gamma^3 N_e(\gamma) \cong  Q_0
 \cases{{\gamma^{3-p}\over [K_{syn{\rm C}}(\gamma_{cool})\gamma_{cool}]}  \; 
,& for 
$\gamma_{min}\lesssim\gamma\leq\gamma_{cool}$  \cr\cr
{\gamma^{2-p}\over K_{syn{\rm C}}(\gamma)} \; , & for
 $\gamma_{cool}\leq \gamma \lesssim\gamma_{max} $   \cr}
\label{gNg}
\end{equation}
where $\gamma_{cool}\equiv \gamma_{cool}(t) $ is found from the
equation $t_{cool}(\gamma_{cool}) = t$. Equation (\ref{gNg}) applies when
$\gamma_{min} < \gamma_{cool}$ and is easily generalized in
the opposite case.  We consider the case where $\gamma_{cool} \ll
\gamma_{\rm KN} \ll
\gamma_{max}$, and 
$\gamma_{max} = \sqrt{ 3e_{max}e/\sigma_{\rm T}B} \cong 4.6\times
10^{10}\sqrt{e_{max}/\Bmug}$ \citep{jag96}, where the parameter
$e_{max}\lesssim 1$.

In the $\delta$-function approximation for the synchrotron and Thomson 
radiation processes, the $\nu F_\nu$ synchrotron radiation spectrum from 
a uniform blob, assumed spherical in the comoving frame, is
\begin{equation}
f_\epsilon^{syn} \simeq \delta^4\; \left( {c\sigma_{\rm T} u_B\over 6\pi 
d_L^2} \right)
\; \gamma_{syn}^3N_e(\gamma_{syn}) \; ,
\label{gdotsyn}
\end{equation}
where $ \gamma_{syn} =\sqrt{(1+z)\epsilon /(\delta\epsilon_B)}$,  
$d_L$ is the luminosity distance, 
$\e = h\nu/m_ec^2$, and $\e_B  = B/4.414\times 10^{13}\,$G. 
The $\nu F_\nu$ Thomson radiation spectrum for an external isotropic 
monochromatic radiation field is
\begin{equation}
f_\epsilon^{\rm T} \simeq \delta^6\; \left( {c\sigma_{\rm T} 
u_*\over 6\pi d_L^2} \right) \; \gamma_{\rm C}^3N_e(\gamma_{\rm C} )
\label{gdotT}
\end{equation}
\citep{dss97,ds01}. Here $\gamma_{\rm C} = \delta^{-1}\sqrt{(1+z)\e/(2\e_*)}$, 
$\epsilon_* = \hat\epsilon_{CMB}(1+z)$ and 
$u_* = \hat u_{CMB}(1+z)^4 $ are the mean photon energy and radiation
energy density, respectively, of the ambient CMBR field in the
stationary frame. An accurate representation of the Compton-scattered
spectrum in the KN regime is given by \citet{gkm01}.

The KN effects on the synchrotron spectrum are observed at $\nu_{{\rm
KN}s1}$(Hz) $\cong$ $ 2\delta \Bmug \gamma_{\rm KN}^2/(1+z)$ $\approx
8\times 10^{16}\delta\Bmug /[\Gamma^2(1+z)^3]$. The reduction of
Compton losses compared to the $-\dot\gamma \propto \gamma^2$ behavior
hardens the electron spectrum until synchrotron losses dominate at
$\gamma > \bar{\gamma}$, resulting again
in a steepening of the synchrotron spectrum in the hard
X-ray domain. 
The KN effects on the
Compton-scattered CMBR spectrum are observed at dimensionless photon
energy $\epsilon_{{\rm KNC}} = h\nu_{\rm KNC}/m_ec^2 
\cong 10^8(\delta/ \Gamma)(1+z)^{-2}$, implying a hardening in the
photon spectrum at photon energy
$E>E_{{\rm KNC}}\cong 50(\delta/\Gamma)(1+z)^{-2}$ TeV.  
We set the normalization $Q_0= (p-2)10^{44}L_{44}{\rm~ergs~s}^{-1}/[(m_ec^2)
(\gamma_{min}^{2-p}-\gamma_{max}^{2-p})]$ and let $p=2.3$,
corresponding to the likely spectral index of particles accelerated by
relativistic shocks \citep{ach01}.

Figs.\ 1a and 1b show two suites of models inspired by the Chandra
data.  In Fig.\ 1a, we assume that $\Gamma = 10$ in a nearby radio
galaxy with $z = 0.15$.
We also assume that the jet axis is inclined
at an angle $\theta = 4/\Gamma = 23^\circ$. The comoving
magnetic field ranges in values from 5 to 50 $\mu$G, and effects of
different values of $\gamma_{cool}$ are illustrated. A field of 44
$\mu$G gives equipartition between the magnetic-field and comoving
CMBR energy densities. The spectral hardening between the optical and
X-ray regimes is apparent.  Note that the use of the $\delta$-function
approximation in the analytic approach enhances spectral features.
 
 The $\Gamma = 10$ jet with $\Bmug = 30$ and $\gamma_{cool} = 10^5$ is
placed at $z = 1$ in Fig.\ 1b, but now with the outer jet oriented at
various angles to the observer. The increasingly dominant Compton
component at small angles results from the different beaming factors
for synchrotron and Compton processes \citep{der95}.  For sufficiently
small values of $\theta$ and $\gamma_{min}$, Thomson-scattered X-ray
CMBR could make a significant or dominant contribution to knot and hot
spot emission, as in knot WK7.8 of the $z = 0.651$ superluminal source PKS
0637-752, with apparent transverse speeds reaching $ \approx 18c$
 \citep{sch00,tav00}.  When observing too close within the
beaming cone, the direct inner jet radiation may however dominate,
which could result in a pattern of $\gamma$-ray flares superimposed on
a significantly weaker but persistent (on time scales of thousands of
years) flux. 

\section{Numerical Model}

Numerical calculations are done using the well-known solution for the
energy distribution $N_{e}(\gamma,t)$ of electrons suffering energy
losses $P(\gamma) \equiv (-\dot\gamma)$, which are contributed mostly
by Compton and synchrotron losses. Calculations are done in the
comoving frame of a source moving relativistically with Lorentz-factor
$\Gamma$, and then the emerging radiation spectra are transformed to
the observer frame.  We neglect possible escape losses of accelerated
particles. This is valid for the large characteristic size of the
X-ray knots of order $R_{knot} \sim 1$-10 kpc or so, because for
$\gamma > 10^4$ the electron cooling time $t_{cool} \lesssim 10^5
\,\rm yr$ in the comoving frame, which is smaller than any reasonable 
escape time from such a large source. Particle escape may
 be important in  much smaller-scale inner jets of blazars 
 \citep{ad01,ad02} or jets of microquasars \citep{aa99}.

Gamma rays with energies $\gtrsim 10^{15} \Gamma^{-1}(1+z)^{-1} \; \rm
 eV$ could be absorbed in $\gamma\gamma$ collisions with the CMB
 photons inside the source. This process provides an injection
 function $Q_{1,\gamma\gamma}(\gamma,t)$ for the first generation
 electrons $(e^+, e^{-})$ of the pair-photon cascade, which we also
 take into account in our calculations. Note however that the cascade
 radiation turns out to be insignificant compared with the radiation
 from the main injection $Q(\gamma,t)$ unless a large fraction of the
 overall power is injected at electron energies $\gtrsim 100$ TeV
 \citep{ad02}.

We consider an electron injection spectrum
\begin{equation}
Q(\gamma,t) \simeq \gamma^{-p} e^{-\gamma/\gamma_{max}} 
\,(1 +t/t_{inj})^{-q} \; 
\label{Qinj} 
\end{equation}
at energies $\gamma_{min} < \gamma $ extending with index $p=2.3$ to
PeV energies, and we set $\gamma_{max}=2\times 10^9$. The injection
function is approximated by $Q \propto (\gamma/\gamma_{cool})^2$ at
$\gamma \lesssim \gamma_{min} = \gamma_{0} \Gamma$ with
$\gamma_{0}=300$.  Note that for electron acceleration by relativistic
shocks, the values of $\gamma_0$ could reach a large fraction of
$(m_{p}/m_{e})\Gamma\sim 2000\Gamma$.  The time profile of the
injection in equation (\ref{Qinj}) treats both stationary (at $t\geq
0$) injection if $t_{inj} \gg t$, as well as gradually declining
injection at $t_{inj} \lesssim t$, when $q>0$.

In Fig.\ 2a we show the synchrotron and Compton fluxes expected from a
relativistic knot with $\Gamma = 10$ in a distant blazar at $z=1$ for
three different times in the knot frame: $t^\prime = 10^4 \,\rm yr$, $5\times
10^4\,\rm yr$, and $3\times 10^5$\,yr\,. If the injection starts
relatively close to the core, then these timescales would effectively
correspond to knots at distances $l= ct^\prime \Gamma \simeq 33 \,\rm kpc$,
$166$\,kpc, and $\sim 1$\,Mpc from the core. Injection is assumed to
be stationary with $L_{44} = 1$, and the magnetic field and the jet
angle to the observer are $B=30\,\rm \mu G$ and $\theta = 11.4^\circ$
respectively.  Fig. 2b shows the fluxes expected at the same epochs
$t^\prime$ from a closer blazar at $z=0.15$, calculated for a smaller
magnetic field $B= 6\,\rm \mu G$ and $\Gamma =5$, and with $\theta =
23^\circ$.  The electron injection here is assumed to decline with
$q=1/2$ and with a characteristic timescale $t_{inj} = 5\times
10^3\,\rm yr$, for initial power $L_{44}= 5$. Fig.\ 2b demonstrates
that the X-ray and optical fluxes may significantly drop while the
radio flux continues to rise with increasing distance, as observed in
3C 273 \citep{sam01}. At the same time, the X-ray fluxes may be
systematically harder than the optical spectra, again in qualitative
agreement with observations of 3C 273.
 
\section{Discussion and Summary}

We have shown that the combined effects of Compton and synchrotron
losses on a power-law electron injection spectrum can produce a
hardening between the optical and X-ray regimes.  This model is in
accord with observations of X-ray spectral hardenings in the knots of
3C 273 \citep{sam01,mar01} and other sources, suggesting that the
X-ray emission is due to a cooling spectrum of electrons accelerated
by a strong shock. In a synchrotron model for the X-rays, the X-ray
profiles are narrower than radio profiles, because the X-rays are
emitted by higher energy electrons and therefore cool more rapidly
than the radio-synchrotron emitting electrons.

The Thomson energy-loss rate must exceed the synchrotron energy
loss-rate to produce X-ray spectral hardenings, so a large fraction of
the energy in nonthermal electrons is radiated as $\gamma$-rays with
energies $\gtrsim 50(\delta/\Gamma)(1+z)^{-2}$ TeV. The
Compton-scattered CMB radiation from the knots and hot spots make
nearly aligned extended jet sources potentially detectable with {\it
GLAST} and the next generation of ground-based air Cherenkov
$\gamma$-ray telescopes such as VERITAS, HESS, and MAGIC. The relative
intensities of the X-ray and $\gamma$-ray fluxes can be used to infer
$\delta$, but the limited spatial resolution of gamma-ray telescopes
will pose difficulties in separating core jet components (excepting
those which are highly variable) from steady extended emission
components. SIRTF will be important to offer greater spectral detail
about the synchrotron component, and to correlate spectral-cooling
breaks in the synchrotron component with the
break in the Thomson component observed with GLAST.
 
The CMB radiation scattered in the Klein-Nishina regime will provide a
source of ultra-high energy gamma-rays that can pair produce in the
diffuse CMB and infrared radiation fields to form pair halos
\citep{acv94}, as well as providing a source of
synchrotron-emitting electrons. The mean-free-path for $\gamma\gamma$
attenuation in the CMB reaches a minimum value of $ \cong 8 {\rm
~kpc}/(1+z)^3$ at observed photon energies $E_\gamma \cong 1$ PeV, and
increases to Mpc scales at lower and higher energies. In the model of
\citet{ad01,ad02}, the injection of energy into the extended jets of
FRII radio galaxies is a consequence of neutral beams composed of
ultra-high energy neutrons and $\gamma$-rays
 ($\sim 10^{13.5}$-$10^{18}\,$eV)
formed through photomeson production in the inner
jet, explaining the colinearity of the inner and extended jets up to
Mpc scales.  The highly collimated neutral-beam energy is deposited in
the intergalactic medium to drive relativistic outflows and form
shocks that accelerate high-energy particles which radiate X-ray
emission observed with Chandra.

\acknowledgments{We thank Hui Li for discussions about the synchrotron 
radiation from high-energy pair cascades, Andrew Wilson for
discussions about the {\it Chandra} data, and the referee for a
useful report. AA appreciates the hospitality and support of the
NRL Gamma and Cosmic Ray (now High Energy Space Environment) Branch
during his visit when this work has been done.  The work of CD is
supported by the Office of Naval Research and NASA grant No.\ DPR
S-13756G.}


\begin{figure}
\epsscale{1.1}
\plottwo{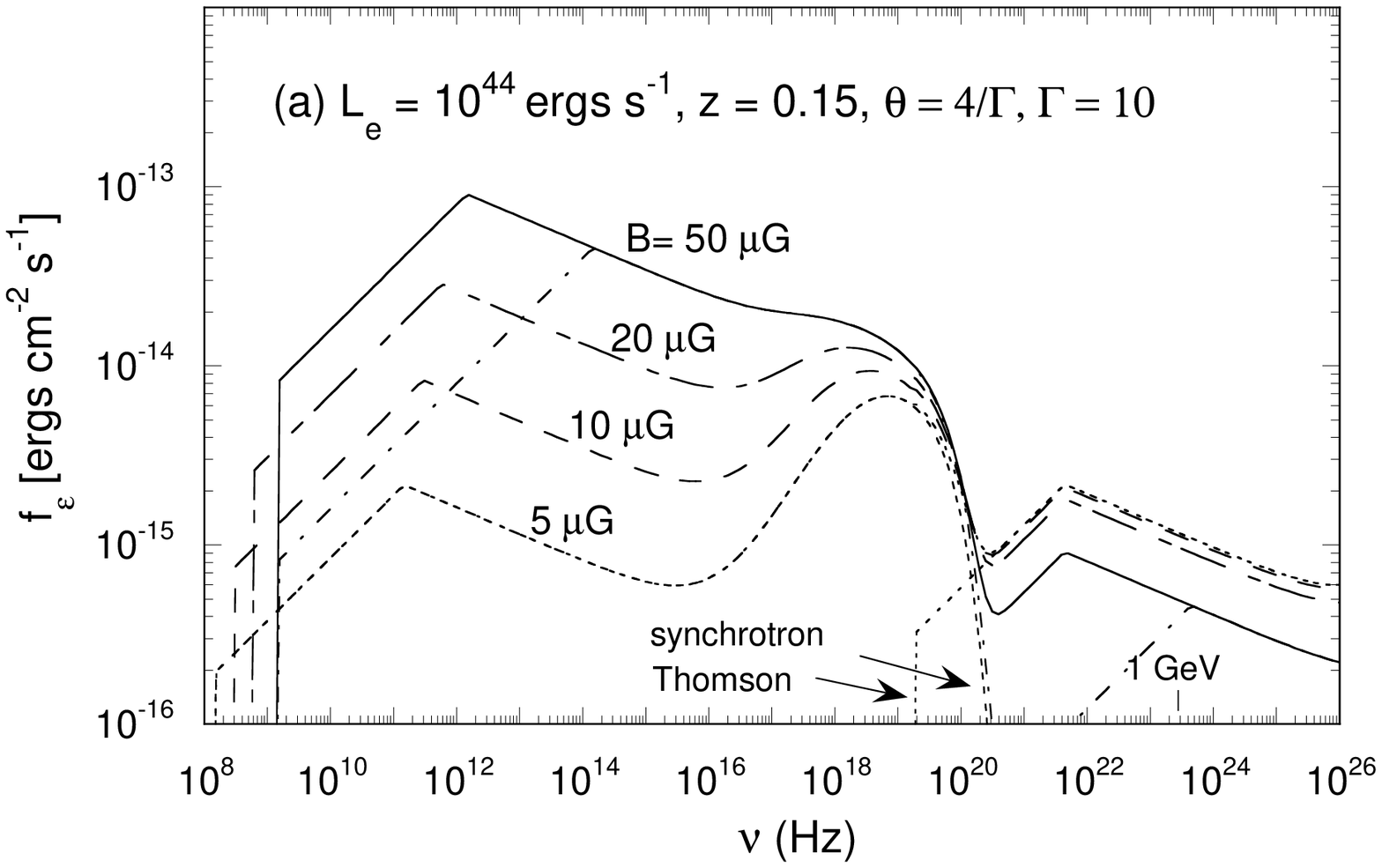}{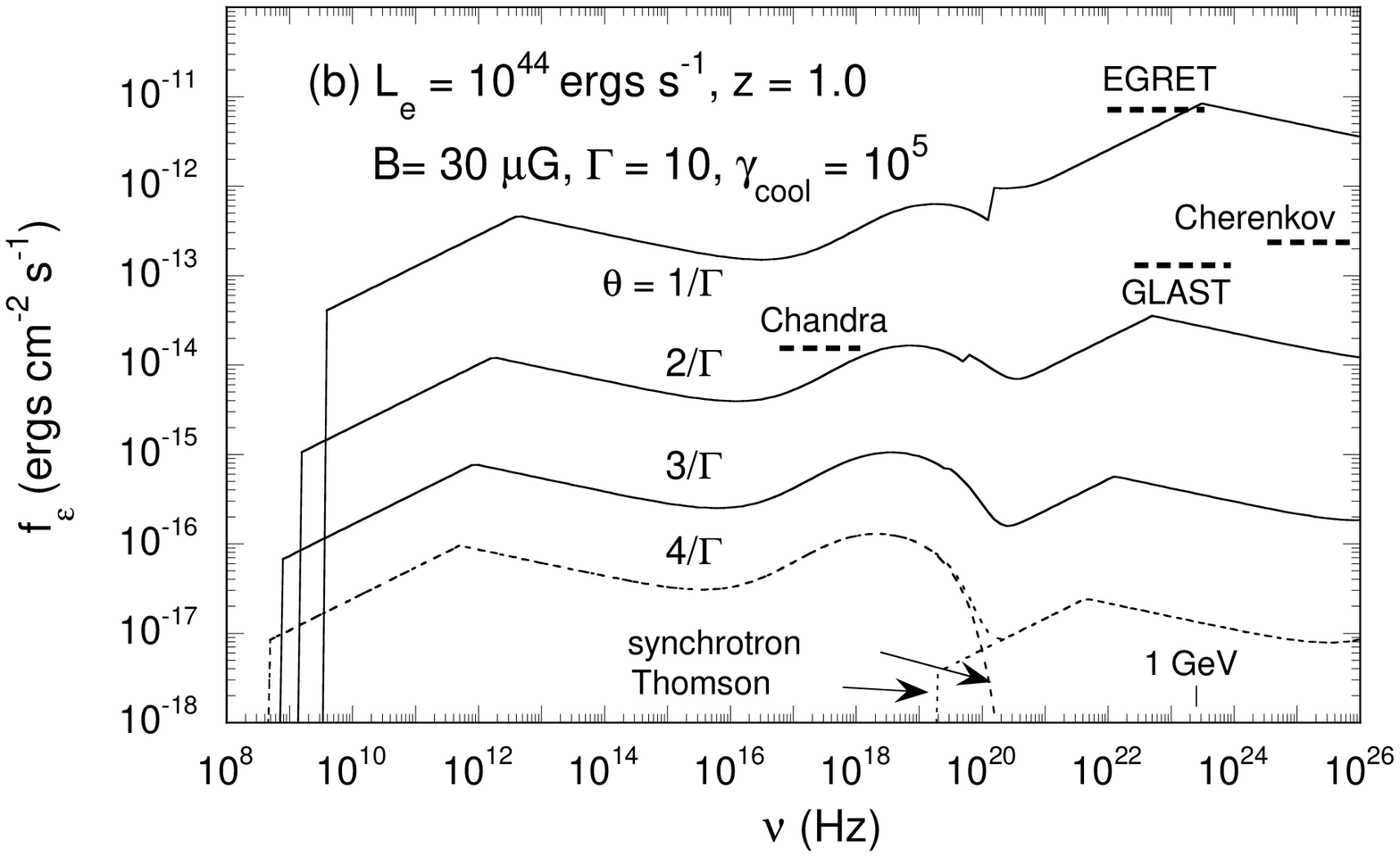}
\caption{Analytic model SEDs radiated by a relativistic 
magnetized knot energized by stationary relativistic power-law
electron injection. The electrons are subject to synchrotron losses
and Compton losses with CMB photons.  (a) The source is at redshift $z
= 0.15$, and the jet is oriented at $\theta = 4/\Gamma$, with bulk
Lorentz factor $\Gamma = 10$ and different comoving magnetic fields as
labeled.  Separate synchrotron and Thomson components are shown in the
$\Bmug = 5$ case. The value of $\gamma_{cool} = 10^5$ except for the
dot-dashed curve, where $\gamma_{cool} = 10^6$.  The injection power
in nonthermal electrons is $L_e = 10^{44}$ ergs s$^{-1}$ ($L_{44} =
1$), $\gamma_{min} = 3000$, and $e_{max} = 0.01$.  (b) Analytic model
SEDs of a source at $z=1$ ($d_L = 2.2\times 10^{28}$ cm) with $\Gamma
= 10$, $\Bmug = 30$, $e_{max} = 0.01$, and $\gamma_{cool} = 10^5$.
Effects of changes in the angle between the jet and observer
directions are shown.  Rough sensitivities for EGRET and GLAST in a
1-year survey, the sensitivity for Chandra from published analyses,
and the anticipated sensitivity of next-generation air Cherenkov
telescopes are shown for comparison.}
\end{figure}

\begin{figure}
\epsscale{1.0}
\plottwo{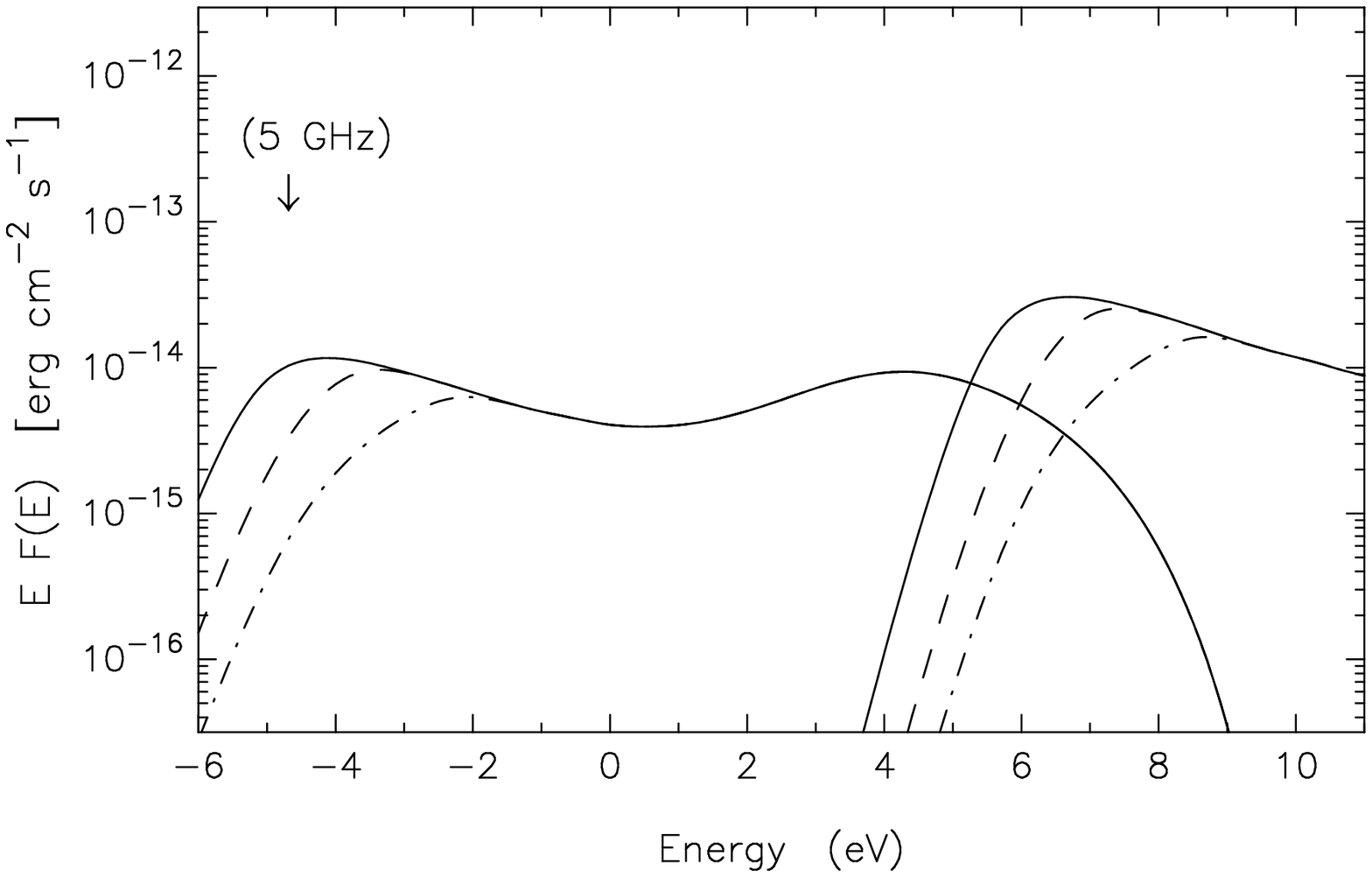}{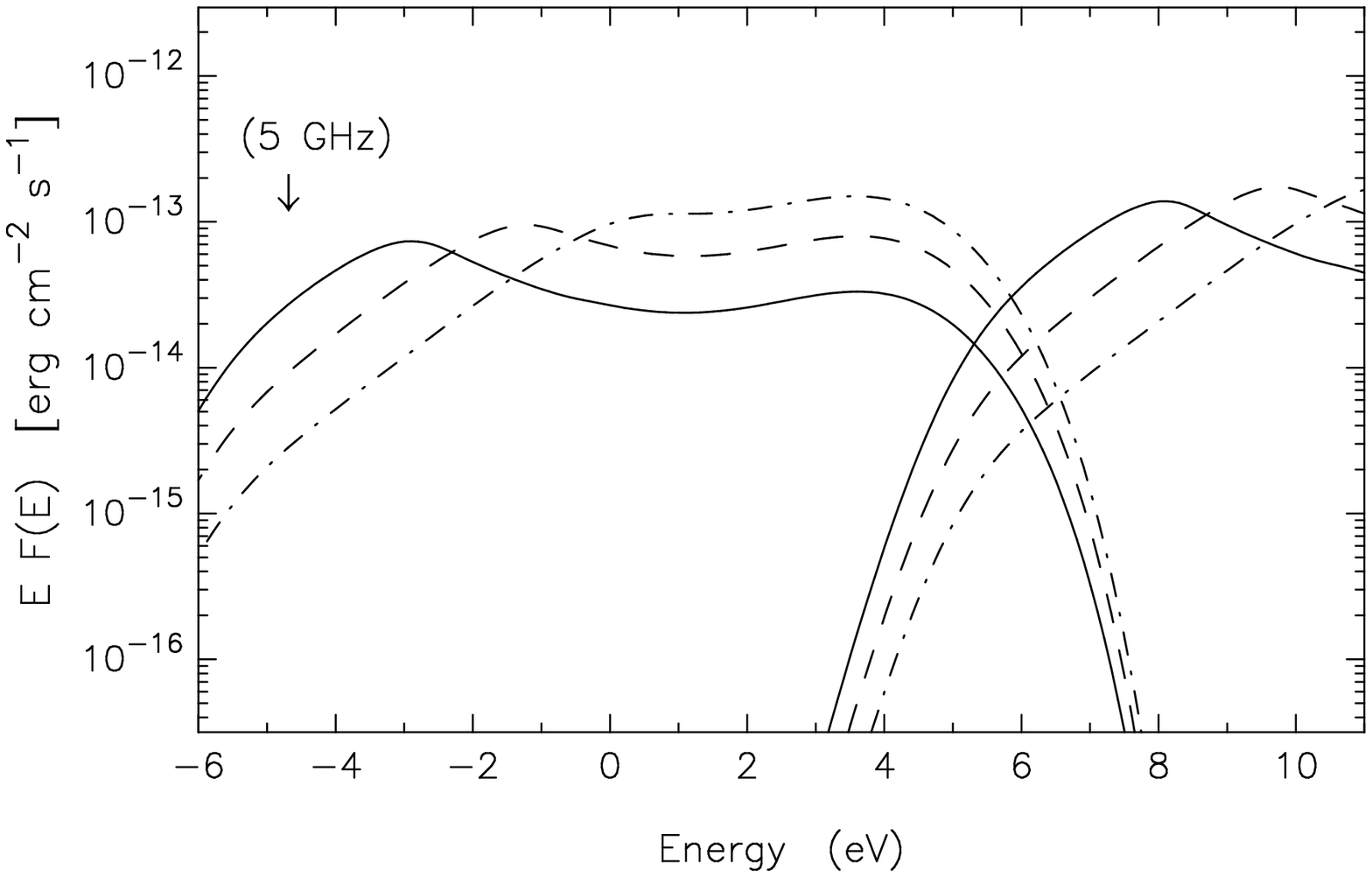}
\caption{Numerical model for fluxes expected from blazar jets at comoving
times $t^\prime=10^4\,\rm yr$ (dot-dashed curves), $5\times 10^4\,\rm yr$
(dashed curves), and $t^\prime= 3\times 10^5\,\rm yr$ (solid curves) after
the start of injection event. (a) Numerical model for fluxes expected
from a distant ($z=1$) blazar jet with stationary injection of
electrons in a relativistic knot for $B=30\,\rm \mu G$, $\Gamma = 10$,
$\theta = 2/\Gamma = 11.4^\circ$, $L=10^{44}\,\rm erg\, s^{-1}$, and
$\gamma_{max} = 8.5\times 10^9$, corresponding to $e_{max} = 1$. (b)
Fluxes from a close blazar ($z=0.15$) for $\Gamma =5$, $B= 6\,\rm \mu
G$, $\theta = 23^\circ$, and $\gamma_{max} = 2\times 10^9$,
corresponding to $e_{max} = 0.01$. A gradually decreasing power of
injection, with $q=0.5$, $t_{inj}=5\times 10^3\,\rm yr$, and an
initial luminosity $L_{44} = 5$ is supposed.}
\end{figure}

\end{document}